\documentclass[graybox]{svmult}

\usepackage{mathptmx}       
\usepackage{helvet}         
\usepackage{courier}        
\usepackage{type1cm}        
\usepackage{makeidx}         
\usepackage{graphicx}        
\usepackage{multicol}        
\usepackage[bottom]{footmisc}

\usepackage{color}
\usepackage{url}
\usepackage[rightcaption]{sidecap}
\usepackage{amsfonts}
\usepackage{amssymb}

\usepackage[normalem]{ulem}  

\begin{document}

\title*{"Strongly interacting matter in magnetic fields": an overview}
\titlerunning{"Strongly interacting matter in magnetic fields": an overview}


\author{Dmitri E. Kharzeev, Karl Landsteiner, Andreas Schmitt, and Ho-Ung Yee}

\institute{Dmitri E.\ Kharzeev \at Department of Physics and Astronomy, Stony
Brook University, Stony Brook, New York 11794-3800, USA, and 
Department of Physics, Brookhaven National Laboratory, Upton, New York
11973-5000, USA, \email{Dmitri.Kharzeev@stonybrook.edu} \and
Karl Landsteiner \at Instituto de F\'{i}sica Te\'{o}rica UAM-CISC, C/Nicol\'{a}s
Cabrera 13-15, Universidad Aut\'{o}noma de Madrid, 28049 Madrid, Spain,
\email{karl.landsteiner@uam.es} \and
Andreas Schmitt\at Institut f\"ur Theoretische Physik, Technische Universit\"at
Wien, 1040 Vienna, Austria, 
\email{aschmitt@hep.itp.tuwien.ac.at} \and
Ho-Ung Yee \at Department of Physics, University of Illinois, Chicago, Illinois
60607, USA, and RIKEN-BNL Research Center, Brookhaven National Laboratory,
Upton, New York
11973-5000, USA,
\email{hyee@uic.edu}}

\maketitle
\abstract{
This is an introduction to the volume of Lecture Notes in Physics on ``Strongly
interacting matter in magnetic fields". The volume combines contributions
written by a number of experts on different aspects of the problem. 
The response of QCD matter to intense magnetic fields has attracted a lot of
interest recently. On the theoretical side, this interest stems from the
possibility to explore the plethora of novel phenomena arising from the
interplay of magnetic field with QCD dynamics. On the experimental side, the
interest is motivated by 
the recent results on the behavior of quark-gluon plasma in a strong magnetic
field created in relativistic heavy ion collisions at RHIC and LHC. The purpose
of
this introduction is to provide a brief overview and a guide to the individual
contributions where these topics are covered in detail.}


\vskip1cm
\section{Introduction}
Electromagnetic probes have proved to be extremely important for understanding
strongly interacting matter -- for example, the discovery of Bjorken scaling
in deep-inelastic scattering (DIS) has allowed to establish quarks as the
constituents of the proton, and has opened the path towards Quantum
Chromodynamics (QCD) and the discovery of  asymptotic freedom. The development
of
QCD has quickly led to the realization that the dynamics of extended field
configurations is crucial for understanding non-perturbative phenomena,
including spontaneous breaking of chiral symmetry and confinement 
that define the properties of our world. 
The challenge of understanding the collective dynamics in QCD calls for the
study of response of strongly interacting matter to intense coherent
electromagnetic fields. Such fields induce a host of interesting phenomena in
QCD matter, and understanding them brings us closer to the ultimate goal of
understanding QCD. 
Some of these phenomena (e.g. Magnetic Catalysis of chiral symmetry breaking
\cite{Shovkovy:2012zn}\footnote{
In this introduction we refer only to the
contributions in the volume, and provide the corresponding arXiv references when
available; many more references can be found in these individual contributions.
\
If you would like to cite one of the contributions on a specific topic, please
refer to them 
directly (instead of citing the entire volume), e.g.
\vskip0.2cm
\noindent
{\it E.~D'Hoker and P.~Kraus, in: ``Strongly interacting matter in magnetic
fields", Eds. D. Kharzeev, K. Landsteiner, A. Schmitt, H.-U. Yee, Lect.~Notes Phys.~871, 467 (2013) [arXiv:1208.1925[hep-ph]].}
\vskip0.2cm
\noindent
If you would like to refer to a broad review representing the collective work of
the authors, you can refer to the entire volume as 
\vskip0.2cm
\noindent
{\it ``Strongly interacting matter in magnetic fields", Eds. D. Kharzeev, K.
Landsteiner, A. Schmitt, H.-U. Yee, Lect.~Notes Phys.~871, 1 (2013) [arXiv:1211.6245[hep-ph]].}}, 
and Inverse Magnetic Catalysis \cite{Preis:2012fh})
exist in a static equilibrium ground state, and affect the phase diagram of  QCD
matter in a magnetic field. 
About one half of this volume addresses such equilibrium phenomena, mostly in
the context of QCD
\cite{Gatto:2012sp,Fraga:2012rr,Chernodub:2012tf,D'Elia:2012tr}, and we give an
overview over this part in Sec.\ \ref{sec:phase}.
The other half addresses mainly anomaly-induced transport phenomena and is
summarized in Sec.\ \ref{sec:cme}. These phenomena include the 
Chiral Magnetic and Chiral Vortical effects (reviewed in
\cite{Zhitnitsky:2012ej,Fukushima:2012vr,Basar:2012gm,Zakharov:2012vv,Gorsky,
Hoyos,Buividovich:2012kq,Yamamoto:2012bi}) which require the existence of a
chirality
imbalance induced by the topological transitions in matter or the presence of
vorticity. 

Experimental access to the study of QCD plasma in very intense magnetic fields
with magnitude $eB \sim m_{\pi}^2$ (or $\sim10^{18}\ \mathrm{G}$) is provided by
collisions of relativistic heavy ions at nonzero impact parameter. 
They create a magnetic field which is (on average) aligned
perpendicular to the reaction plane. 
Somewhat weaker magnetic fields $\sim 10^{15}\ \mathrm{G}$ exist on the surface
of magnetars. They are possibly much larger in the interior of the star, 
where they may affect the properties of cold dense quark matter as reviewed in
\cite{Ferrer:2012wa}.

\section{Chiral magnetic effect and anomaly-induced transport}
\label{sec:cme}

The Chiral Magnetic Effect (CME) is the phenomenon of electric charge separation
along an external magnetic field that is induced by a chirality imbalance. 
In QCD matter, the source of chirality imbalance are the transitions between the
topologically distinct states 
-- the index theorem and the axial anomaly relate the resulting change in
topological number to the chirality of fermion zero modes. The CME is thus a
topological effect: it 
results from an interplay of topology of the zero mode of a charged fermion in
an external Abelian magnetic field, and of the non-Abelian topology of 
the gluon field configurations. Because of this, the CME current is
topologically 
protected (not sensitive to local perturbations) and non-dissipative.

At weak coupling, the quasi-particle picture is appropriate, and allows to
understand the phenomenon in a very simple 
and intuitive way, as we now explain. An external magnetic field aligns the
spins of the positive and negative fermions at the lowest Landau level in 
opposite directions (only the lowest Landau level matters for the CME, as the
contributions of all excited levels cancel out -- see \cite{Fukushima:2012vr}
for details). Therefore, the electric charge, chirality and momentum of the
fermion are
correlated -- for example, a positively charged right-handed fermion propagates
along the direction of magnetic field, and a negative right-handed fermion
propagates in the opposite direction. This creates an electric current, which
however is 
usually compensated by the left-handed fermions that propagate in the opposite
direction. 

Let us however imagine that the fermions, apart from the Abelian $U(1)$ charge,
are also charged under a non-Abelian group -- for us, the most important example
is provided by quarks, which carry both electric and color charges. The
non-Abelian gauge theories possess a rich spectrum of topological solutions, and
the axial anomaly links the topology of gauge fields to 
the chirality of fermions. Therefore, in a topologically non-trivial non-Abelian
background, the numbers of left- and right-handed fermions will in general 
differ -- because of this, their contributions to the electric current will no
longer cancel. As a result, an external magnetic field will induce an electric
current along its direction -- an effect that is absent in Maxwell
electrodynamics. 

The absence of CME in conventional electrodynamics follows already from 
symmetry considerations -- the magnetic field is a (parity-even)
pseudo-vector, and the electric current is a (parity-odd) vector. Therefore, CME
signals the violation of parity -- indeed, as we discussed above, its presence
requires the asymmetry between the left and right fermions.

It is well known that there are no perturbative corrections to the axial
anomaly, so the CME expressions for the electric current and electric dipole
moment 
are exact (at the operator level). Moreover, because the origin of CME is
topological, it appears that
the CME current at zero frequency remains the same even in the limit of strong
coupling, that is accessible theoretically through the holographic
correspondence -- see  \cite{Hoyos,Landsteiner:2012kd}. 
A phenomenon similar to CME arises when instead of a magnetic field there is an
angular momentum  (vorticity) present -- this is the so-called Chiral Vortical
Effect (CVE). The CVE is also caused by the quantum anomaly, but by the
gravitational one  \cite{Landsteiner:2012kd}. In a holographic setup, this is
described through a mixed gauge-gravitational Chern-Simons term.

In the absence of external charge, magnetic field in the framework of AdS/CFT
correspondence induces an RG flow to an infrared $AdS_2\times \mathbb{R}^2$
geometry \cite{D'Hoker:2012ih}. This theory is the holographic description of
the dimensionally reduced 2d conformal field theory describing the strongly
coupled fermions on lowest Landau levels. In general, the dimensional reduction
proves very beneficial to treating the CME and related phenomena -- see
\cite{Basar:2012gm} for review.
In particular, in the limit of strong magnetic field one can construct an
explicit solution describing the QCD instanton in magnetic background
\cite{Basar:2012gm}-- since the instanton induces an asymmetry between left- and
right-handed fermions, this solution has been found to possess electric dipole
moment, in accord with CME expectations. It is of great interest to investigate
the dynamics of CME by considering the decay of topological objects in magnetic
backgrounds -- see \cite{Gorsky} for a review.

The persistence of CME at strong coupling and small frequencies makes the
hydrodynamical description of the effect possible, as reviewed in
\cite{Zakharov:2012vv}. The quantum anomalies in general have been found to
modify hydrodynamics in a significant way. This has a profound importance for
transport, as 
the anomalies make it possible to transport currents without dissipation -- this
follows from the $P$-odd and $T$-even nature of the corresponding transport
coefficients. The existence of CME and CVE in hydrodynamics is interesting also
for the following reason -- usually, in the framework of quantum field theory
one thinks about quantum anomalies as of UV phenomena arising from the
regularization of loop diagrams. However, we now see that the anomalies also
modify the large distance, low frequency, response of relativistic fluids. This
is because the anomalies link the chirality of fermion zero modes to the global
topology of gauge
fields.

The CME can be studied numerically from first principles on Euclidean
space-time lattices, see \cite{Buividovich:2012kq,Yamamoto:2012bi} for reviews.
On the lattice, one can measure the fluctuations of electric charge asymmetry
induced by the dynamical topological fluctuations in QCD vacuum and plasma in
a magnetic background \cite{Buividovich:2012kq}. Alternatively, one can
introduce
the chiral chemical potential, and measure the CME current explicitly, testing 
the relation between the current, magnetic field, and the chiral chemical
potential -- this approach is reviewed in \cite{Yamamoto:2012bi}. Note that the
chiral chemical potential 
(unlike the baryon one) does not lead to the determinant ``sign problem" and
thus does not prevent one from 
performing lattice QCD simulations.

From the experimental viewpoint, CME makes it possible (at least in principle)
to observe directly the fluctuations of topological charge in heavy ion
collisions -- indeed, these fluctuations in magnetic field induce the asymmetry
of electric charge distributions with respect to the reaction plane. Such a
study has to
carefully separate the CME effects from all possible backgrounds, as reviewed in
\cite{Bzdak:2012ia}. One of the CME tests discussed in \cite{Bzdak:2012ia} is
the collision of two Uranium ions
where the deformation of the Uranium nucleus allows to separate the CME from the
backgrounds, as we now explain. 

 The main idea behind the $UU$ measurement is the
following: all possible backgrounds to CME should, on symmetry grounds, be
proportional to the elliptic flow of hadrons. The elliptic flow stems from the
ellipticity of the initial fireball produced in heavy ion collisions, and so it
exists only in non-central collisions, just like the magnetic field that drives
the
CME. This complicates the  separation of CME from possible background effects
\cite{Bzdak:2012ia}. However, since the Uranium nucleus is strongly deformed,
even most central $UU$ collisions (where there is no magnetic field) produce
a deformed fireball and hence a sizable elliptical flow of hadrons. Thus, if
the charge separation persists in most central $UU$ collisions, the observed
effect is most likely due to background, and vice versa. Very recently, the 
RHIC result on $UU$ collisions has been presented at the Quark Matter 2012
Conference by the STAR Collaboration. The presented result
indicates that the signal vanishes in the absence of a magnetic field, 
providing a strong support for the CME interpretation.

\section{Phase structure in a magnetic field}
\label{sec:phase}

Equilibrium properties of matter can be changed significantly by the presence of
a background magnetic field. This is true even for 
non-interacting matter. Consider for example a Fermi sphere of charged,
non-interacting fermions, say at zero temperature, that is subject to a static
and 
homogeneous magnetic field. Instead of a continuous spectrum with respect to all
three momentum directions, the system will develop discrete energy levels
with respect to the momentum directions transverse to the magnetic field. Only
the longitudinal momentum remains continuous. If the magnetic field is large
enough,
i.e. of the order of the Fermi momentum squared, the discretization of
the energy levels, called Landau levels, will have an important effect on the 
properties of the system. For instance, upon increasing the magnetic field, the
distance between the energy levels increases and as these levels ``pass''
the given Fermi energy, the observable quantities such as the number density
change.
Eventually, for sufficiently 
large magnetic fields, all fermions reside in the lowest Landau level,
where only one spin polarization is allowed. The system has become fully 
polarized.

A more challenging question is how the equilibrium properties of {\it strongly
interacting} matter are affected by a magnetic field. One might for instance ask
whether there are still Landau levels or whether this description becomes
incorrect. One might also wonder about the phase transitions induced by magnetic
field --  
do we enter genuinely different phases upon increasing a background magnetic
field in a given, strongly-interacting, system? If yes, what are the order 
parameters,  what is the order of these phase transitions, and what are the
critical magnetic fields? Or, ultimately, what is the phase diagram of a 
given system with one of the axes corresponding to the magnetic field? A
well-studied example from condensed-matter physics is the phase diagram of
Helium-3.
Due to the nontrivial structure of the order parameter with respect to spin and
orbital angular momentum, Helium-3 has more than one superfluid phase, 
and an externally applied magnetic field can induce phase transitions between
different superfluid phases. 

\subsection{Phases of QCD in a magnetic field}


In these Lecture Notes, the term ``strongly interacting matter" mostly refers to
the 
matter governed by 
QCD, although some chapters address questions from ``ordinary'' condensed-matter
systems \cite{D'Hoker:2012ih,Albash:2012ht,Gubankova:LNP}, see also the
discussion about graphene in 
Sec.\ 3.3 of Ref.\ \cite{Shovkovy:2012zn}. Usually, ``the phase diagram of QCD''
is drawn in the plane spanned by the temperature $T$ and the baryon chemical
potential 
$\mu$. But various additional directions, i.e., higher-dimensional versions of
the phase diagram, are of interest as well. First, one can imagine to change the
parameters of QCD by hand, for example by adding an axis for the quark mass(es)
or for the number of colors $N_c$ to the phase diagram. The purpose of such a
theoretical manipulation can be to enter a more tractable regime and/or to put
the 
QCD phase structure into a wider context, with the ultimate goal to
understand better the phase structure of real-world QCD. Second, there are 
external parameters of direct phenomenological interest that reflect the
characteristics of matter, such as the baryon and isospin chemical potentials
that become important in the context of neutron stars. A (uniform) magnetic
field is both: it is of phenomenological interest but also a
theoretically 
useful ``knob'' by which interesting and rich physics can be introduced which
may help us to deepen our understanding of strongly 
interacting systems. (This volume mostly
addresses the effects of a magnetic field from ordinary $U(1)$
electromagnetism; the 
effects of chromomagnetic fields are briefly discussed in
\cite{D'Elia:2012tr}.) 

Phase transitions in QCD can be expected to
occur at energy scales comparable to the QCD scale $\Lambda_{\rm QCD}\sim
200\,{\rm
MeV}$.
As a consequence, we are interested in magnetic field strengths of the order of
$B\sim (200\,{\rm MeV})^2\simeq 2\times 10^{18}\,{\rm G}$. As mentioned above,
such strong 
magnetic fields indeed exist in non-central heavy-ion collisions
\cite{Bzdak:2012ia} (at least temporarily), and possibly in the interior of 
compact stars, then called magnetars \cite{Ferrer:2012wa}. 
These two instances are (together with the early universe) the only systems with
which we can reach out ``experimentally'' also into the $\mu$-$T$ plane of the
QCD
phase 
diagram
and thus probe the phase structure of QCD. Most notably, we are interested in
the nonperturbative phenomena of confinement and chiral symmetry breaking. 
Both heavy-ion collisions and compact stars are expected to ``live'' in a region
of the phase diagram where the transitions to deconfined and/or 
chirally restored matter occur and thus it is relevant to discuss the location
and the nature of these transitions in magnetized QCD matter.

The various chapters addressing the phase structure of QCD in magnetic fields
\cite{Shovkovy:2012zn,Preis:2012fh,Gatto:2012sp,Fraga:2012rr,
Chernodub:2012tf,D'Elia:2012tr,Ferrer:2012wa,Bergman:2012na} make
use of different theoretical tools. At asymptotically large energies, methods
from perturbative QCD can be applied, see parts of
\cite{Shovkovy:2012zn,Ferrer:2012wa}. For moderate energies, however,
first-principle QCD calculations can only be done on the lattice and are
restricted 
to vanishing (vector) chemical potentials \cite{D'Elia:2012tr}. These results
from QCD
are complemented by model calculations, which may differ (and, judging from the
results 
presented here, {\it do} differ) in important aspects from actual QCD.
Nevertheless, they may be useful to get an idea about some of the 
physical mechanisms behind the phase structure. The models discussed here are
the Nambu--Jona-Lasinio (NJL) model, including different variants with 
respect to the interaction terms and extensions to incorporate confinement
(PNJL) \cite{Shovkovy:2012zn,Preis:2012fh,Gatto:2012sp,Chernodub:2012tf}, 
the MIT bag model \cite{Fraga:2012rr}, and a quark-meson model
\cite{Gatto:2012sp,Fraga:2012rr}.
Additionally, we employ the gauge/gravity duality
\cite{Bergman:2012na,Preis:2012fh} which provides us with a reliable tool for
the physics in the 
strongly coupled limit, albeit for theories that differ more or less --
depending on the
model at hand --  from  real QCD. 

The holographic calculations presented here make use of two different setups.
Firstly, based
on the original AdS/CFT correspondence,
a background geometry given by D3-branes is discussed, where D7-branes,
corresponding to fundamental (as opposed to adjoint) degrees of freedom, are
introduced as 
probe branes \cite{Bergman:2012na}. And secondly, we use the Sakai-Sugimoto
model, where ``chiral'' D8-branes are embedded into a background of D4-branes, 
which includes a compact extra dimension that serves to break supersymmetry
completely \cite{Preis:2012fh,Bergman:2012na}. 

The picture that emerges from these studies is, concerning QCD, a preliminary
one at best, and many questions are still open -- and, in fact,
are 
raised by these studies. 
Nevertheless, let us try to summarize the current picture. Before one addresses
the question of how a magnetic field affects a hot and dense medium, one might
ask 
whether and how the vacuum changes in a magnetic field. A very important part of
the
answer to this question -- since it is of very general nature and thus not 
only relevant for QCD -- is ``magnetic catalysis'', which is reviewed in great
detail in Ref.\ \cite{Shovkovy:2012zn}. This effect has been confirmed in
numerous
model calculations, as well as in QED-like theories and -- at least at zero
temperature -- in QCD lattice calculations. In simple words, it says that a
magnetic
field favors chiral symmetry breaking. A more precise version, for instance
employing the mean-field approximation of the NJL model, is as follows. For
$\mu=T=0$ and vanishing magnetic field, there is a critical coupling strength
above which a chiral condensate forms, i.e., there
is a phase transition from the chirally restored to the chirally broken phase as
one increases the (attractive) coupling of the four-fermion interaction. 
In the presence of a background magnetic field, however, there is a chiral
condensate for {\it arbitrarily small} coupling. Thus, the magnetic field has a
profound
qualitative effect on chiral symmetry breaking. One way to understand the
physics behind magnetic catalysis is the analogy to BCS Cooper pairing. 
In both cases, the 
dynamics of the system becomes effectively 1+1 dimensional at weak coupling, in
the case of Cooper pairing because of the presence of the Fermi surface, in the
case of 
chiral condensation because of the magnetic field. As a consequence, an infrared
divergence occurs which is cured by a nonvanishing mass gap
that shows precisely the same exponential, nonperturbative behavior in both
cases.    

Another possible effect of an ultra-strong magnetic field on the QCD vacuum is
the condensation of $\rho$ mesons \cite{Chernodub:2012tf}. In a weak-coupling
picture, 
this condensation is suggested from the Landau-level structure of spin-1 bosons.
Since $\rho$ mesons are electrically charged, their condensation implies 
electric superconductivity. The details of this interesting idea are reviewed in
Ref.\ \cite{Chernodub:2012tf}.

How does magnetic catalysis manifest itself in the QCD phase diagram? A
straightforward, but, as we now know, too naive, expectation is that the phase
space region
of the chirally broken phase should get larger with the magnetic field. The
current picture is more complicated and can be summarized as follows.
\begin{itemize}
\item {\it Hot medium, $\mu=0$.} 
Lattice studies with physical quark masses suggest that the (pseudo-)critical
temperature $T_c$ for the chiral crossover  {\it decreases} 
\cite{D'Elia:2012tr}, 
whereas the results of all above mentioned model calculations (as well as
lattice
results for unphysically large quark masses \cite{D'Elia:2012tr})
show a monotonically {\it increasing} $T_c$. While it is clear that none of the
models can capture 
all features of QCD, the physical mechanism behind the lattice result is still
under discussion, see for instance Sec.\ 3.2 in Ref.\ \cite{Shovkovy:2012zn}. 
There are also slight differences in the behavior
of $T_c$ in the different models: for instance, while $T_c$ saturates 
at a finite value for asymptotically large magnetic fields in the holographic
Sakai-Sugimoto model \cite{Preis:2012fh,Bergman:2012na}, no such saturation can
be 
seen in the NJL-like models \cite{Preis:2012fh,Gatto:2012sp}. 

\item {\it Dense medium, $T=0$.} In this case, there are currently no lattice
results due to the sign problem. In the model calculations we see
an interesting nontrivial effect, termed ``inverse magnetic catalysis''
\cite{Preis:2012fh}. At strong coupling, the 
critical chemical potential can {\it decrease} with the magnetic field, which is
in apparent conflict with magnetic catalysis. In contrast to the 
$\mu=0$ result on the lattice, a physical explanation for this behavior is
known. It can be traced back to the cost in free energy that has to be paid for
chiral 
condensation at nonzero $\mu$. Crucially, this cost is not independent of $B$
and competes with the gain from condensation which, due to magnetic catalysis,
increases 
with $B$.

\end{itemize}

We know that the QCD phase structure at large densities
can be very rich due to various
color-superconducting phases. Model calculations at nonzero
$B$ including two-flavor color superconductivity seem to confirm the 
effect of inverse magnetic catalysis. From first
principles we know that at asymptotically large densities, the ground state of
three-flavor
quark matter is the color-flavor locked (CFL) phase.  The CFL order parameter is
invariant under 
a certain combination of generators of the color and electromagnetic gauge
groups, resulting in a massless gauge boson (which is predominantly the original
photon, 
with a small admixture from one of the gluons). As a consequence, the CFL phase
is a color superconductor (all gluons become massive), but not an
electromagnetic 
superconductor. 
An ordinary magnetic field can thus penetrate this phase. Depending on the
strength of the magnetic field, certain variants of the CFL phase 
become favored. These phases as well as their possible astrophysical relevance
are discussed in  \cite{Ferrer:2012wa}.

For the deconfinement crossover in a magnetic field, the results on the lattice
are similar to the
chiral crossover \cite{D'Elia:2012tr}. Again, with 
physical quark masses the temperature for the crossover seems to decrease. This
behavior is not reproduced by the PNJL model, where 
confinement is mimicked by including the expectation value of the Polyakov loop
by hand. Depending on the details of the interactions within the model, 
the deconfinement and chiral phase transitions do or do not split in a magnetic
field, but in any case the critical temperature of both transitions seems to
increase 
monotonically with the magnetic field \cite{Gatto:2012sp}. This is also the case
in a quark-meson model \cite{Fraga:2012rr}. 
Within the MIT bag model, however, with the confined phase simply being
modelled 
by a noninteracting pion gas,
the deconfinement transition decreases with the magnetic field, in qualitative
agreement with the lattice results \cite{Fraga:2012rr}. These studies show
that the mechanism behind the behavior of the deconfinement transition -- just
like for the chiral transition --  in a magnetic field is not yet understood, 
and further analyses that clarify this picture are necessary. 

\subsection{Condensed matter systems in a magnetic field via AdS/CFT}

The gauge-gravity or AdS/CFT correspondence plays an important role
in the above mentioned studies of the phase structure
of QCD. This is quite natural since in general the gauge-gravity correspondence
relates a non-Abelian gauge theory at strong coupling
and large $N$ to a weakly coupled (super)gravity background with asymptotic AdS
boundary conditions. The idea that strongly coupled
phases of quantum field theories can be described by a gravity dual is however
more general than that. In particular it has been applied
to ``tabletop`` laboratory condensed matter systems that inherently involve
strongly correlated electrons and are conjectured
to be governed by an underlying quantum critical phase such as high Tc
superconductors or the so called strange metals. Many condensed-matter systems
are supposed
to undergo a quantum phase transition at zero temperature upon varying a
parameter such as the doping in a high-$T_c$ superconductor, the pressure or the
applied magnetic field. At the phase transition the system is at  a quantum
critical point which often turns out to be
in a strongly coupled regime with an emergent Lorentz (or more generally
Lifshytz) invariance. In the vicinity of the quantum critical point, e.g. at
finite temperature the dynamics of the system is still governed by the degrees
of
freedom relevant at the critical point.  One might hope
that such strongly coupled quantum critical points can be described by a
gravity dual just as QCD in its strongly coupled regime
might be described by e.g. the Sakai-Sugimoto model. At the very least one can
expect that the gauge-gravity duality helps to develop
interesting toy models of quantum critical points that can lead to a better
understanding of qualitative or even quantitative behavior of 
condensed matter systems whose theoretical description is otherwise rather
elusive due to their strongly coupled nature. 

A central role is played by charged asymptotically AdS black holes. Due to the
underlying scaling symmetry two regimes can naturally
be distinguished, one in which the temperature T is much larger than the
chemical potential $\mu$. This is the hydrodynamic regime and the
gauge gravity duality allows to compute transport coefficients such as
viscosities or conductivities. The other regime of interest
is the opposite with $\mu \gg T$. Here the black hole is near extremality. At
precisely $T=0$ the black hole horizon becomes degenerate
with profound implications for the dual field theory physics. It turns out that
the entropy of an extremal charged AdS black hole is macroscopically large,
scaling with some positive power of the number of the underlying microscopic
degrees of freedom $N$. This feature
is generally interpreted as pointing towards an inherent instability of such
black holes. Indeed many such instabilities are known to 
arise upon adding additional fields, e.g. scalars (even uncharged ones) tend to
condense near extremality leading to a superconducting
phase transition. The charge that was hidden behind the horizon is pulled
outside 
and the scalar field forms the (symmetry breaking) condensate.
If one replaces the scalar with a fermion a similar transition occurs where
the black hole is replaced by a geometry without horizon and the charge is
carried by the fermions forming an "electron star" in 
asymptotically AdS.  The field theory interpretation of this transition is one
between a "fractionalized" phase  and a "mesonic" phase.
This is in one-to-one correspondence to the deconfined/confined phases of
non-Abelian gauge theory, only that in condensed matter
it is often the mesonic or confined phase that appears as the more fundamental
one. Real world fractionalization occurs for example
in the separation of spin and charge degrees of freedom of electrons or the
appearance of quasiparticles carrying fractions
of the electron's charge. Such gravity duals of fractionalized/mesonic phases in
the presence of a magnetic field 
are the subject of \cite{Albash:2012ht}. A crucial role is played by the
presence of a dilaton which allows to find a rich structure of solutions
including partially fractionalized phases. 

The quantum critical behavior of four dimensional field theories described by a
gravity dual including a $U(1)$ gauge field with
an additional Chern-Simons term in a magnetic field is the subject of
\cite{D'Hoker:2012ih}. The authors find a quantum phase transition
for large enough magnetic field beyond which the the charge is completely
expelled outside the horizon and the background geometry
has zero entropy. 

Another important theme in the condensed matter applications of the gauge
gravity correspondence is the spectrum of fermions in
asymptotically AdS black holes. Fermions in the gravity dual obeying a Dirac
equation dual to gauge invariant fermionic operators in
the field theory. One can study the holographic fermionic two point function and
in particular look out for poles that identify the presence
of Fermi surfaces. It is indeed well known by now that such probe fermions can
show behavior consistent with Landau's theory
of Fermi liquids but also more exotic possibilities such as marginal Fermi
liquids in which the residue of the corresponding pole vanishes or completely
non-Landau Fermi liquid behavior are realized. The Fermi level structure of
such
probe fermions in a four dimensional dyonic, i.e. including a magnetic field,
asymptotically  AdS black hole is the subject of \cite{Gubankova:LNP}. For
strong magnetic field the Fermi surface vanishes and the authors associate this
with a metal to strange metal phase transition.
 
 To summarize, this volume of Lecture Notes presents a review of the current
research of strongly interacting matter in magnetic fields. 
 Most of the applications considered here 
 concern QCD matter, but a number of important cases from condensed matter
physics is considered as well. 
 The focus of the volume is on the theoretical results; however these results 
 have a direct significance for experiment, in particular for heavy ion
collisions that are currently under intense 
 study at RHIC and LHC. While most of the contributions in this volume reflect
the work done very recently, 
 the field is evolving so rapidly that we expect to see a significant progress
already in the near future. 
 Because of this, we do not expect this volume to express a ``final word" in any
sense; instead, we view it as a snapshot of
 the exciting work that is being done right now. A large number of open
questions has emerged as a result of this work; some of them have 
 been mentioned in this brief overview, but we encourage the reader to read the
individual contributions for an in-depth exposure. 
 We hope that this volume will convince the reader that strongly interacting
matter in a magnetic field is a rich and vibrant research area, and  
 many more discoveries and surprises can be fully expected.

 \begin{acknowledgement}
We
would like to 
express our gratitude to all authors of these Lecture Notes for their
contributions.
This overview was finalized during the workshop "QCD in strong magnetic
fields" at the European Centre for Theoretical Studies in Nuclear Physics and
Related Areas (ECT*). We would like to thank ECT* for its kind hospitality and
the organizers for a very interesting and stimulating workshop.  The work of
D.K. is supported in part by the US Department of Energy under Contracts
DE-AC02-98CH10886 and DE-FG-88ER41723.
The work of K.L. is supported by Plan Nacional de Altas Energ\'\i as
FPA2009-07890, 
Consolider Ingenio 2010 CPAN CSD2007-00042 and HEP-HACOS S2009/ESP-2473.

\end{acknowledgement}

\bibliographystyle{unsrt}
\bibliography{refs}

\begin{thebibliography}{10}

\bibitem{Shovkovy:2012zn}
Igor~A. Shovkovy.
\newblock {"Magnetic Catalysis: A Review"}.
\newblock {\em Lect.~Notes Phys.}, 871:13, 2013.
\newblock arXiv:1207.5081[hep-ph].

\bibitem{Preis:2012fh}
Florian Preis, Anton Rebhan, and Andreas Schmitt.
\newblock {"Inverse magnetic catalysis in field theory and gauge-gravity
  duality"}.
\newblock {\em Lect.~Notes Phys.}, 871:49, 2013.
\newblock arXiv:1208.0536[hep-ph].

\bibitem{Gatto:2012sp}
Raoul Gatto and Marco Ruggieri.
\newblock {"Quark Matter in a Strong Magnetic Background"}.
\newblock {\em Lect.~Notes Phys.}, 871:85, 2013.
\newblock arXiv:1207.3190[hep-ph].

\bibitem{Fraga:2012rr}
Eduardo~S. Fraga.
\newblock {"Thermal chiral and deconfining transitions in the presence of a
  magnetic background"}.
\newblock {\em Lect.~Notes Phys.}, 871:119, 2013.
\newblock arXiv:1208.0917[hep-ph].

\bibitem{Chernodub:2012tf}
M.N. Chernodub.
\newblock {"Electromagnetic superconductivity of vacuum induced by strong
  magnetic field"}.
\newblock {\em Lect.~Notes Phys.}, 871:141, 2013.
\newblock arXiv:1208.5025[hep-ph].

\bibitem{D'Elia:2012tr}
Massimo D'Elia.
\newblock {"Lattice QCD Simulations in External Background Fields"}.
\newblock {\em Lect.~Notes Phys.}, 871:179, 2013.
\newblock arXiv:1209.0374[hep-lat].

\bibitem{Zhitnitsky:2012ej}
Ariel~R. Zhitnitsky.
\newblock {"P odd fluctuations and Long Range Order in Heavy Ion Collisions.
  Deformed QCD as a Toy Model"}.
\newblock {\em Lect.~Notes Phys.}, 871:207, 2013.
\newblock arXiv:1208.2697[hep-ph].

\bibitem{Fukushima:2012vr}
Kenji Fukushima.
\newblock {"Views of the Chiral Magnetic Effect"}.
\newblock {\em Lect.~Notes Phys.}, 871:239, 2013.
\newblock arXiv:1209.5064[hep-ph].

\bibitem{Basar:2012gm}
G{\"{o}}k\c{c}e Ba\c{s}ar and Gerald~V. Dunne.
\newblock {"The Chiral Magnetic Effect and Axial Anomalies"}.
\newblock {\em Lect.~Notes Phys.}, 871:259, 2013.
\newblock arXiv:1207.4199[hep-th].

\bibitem{Zakharov:2012vv}
Valentin~I. Zakharov.
\newblock {"Chiral Magnetic Effect in Hydrodynamic Approximation"}.
\newblock {\em Lect.~Notes Phys.}, 871:293, 2013.
\newblock arXiv:1210.2186[hep-ph].

\bibitem{Gorsky}
A.~Gorsky.
\newblock {"Remarks on Decay of Defects with Internal Degrees of Freedom"}.
\newblock {\em Lect.~Notes Phys.}, 871:329, 2013.

\bibitem{Hoyos}
Carlos Hoyos, Tatsuma Nishioka, and Andy O'Bannon.
\newblock {"A Chiral Magnetic Effect from AdS/CFT with Flavor"}.
\newblock {\em Lect.~Notes Phys.}, 871:339, 2013.

\bibitem{Buividovich:2012kq}
P.V. Buividovich, M.I. Polikarpov, and O.V. Teryaev.
\newblock {"Lattice studies of magnetic phenomena in heavy-ion collisions"}.
\newblock {\em Lect.~Notes Phys.}, 871:375, 2013.
\newblock arXiv:1211.3014[hep-ph].

\bibitem{Yamamoto:2012bi}
Arata Yamamoto.
\newblock {"Chiral Magnetic Effect on the Lattice"}.
\newblock {\em Lect.~Notes Phys.}, 871:385, 2013.
\newblock arXiv:1207.0375[hep-lat].

\bibitem{Ferrer:2012wa}
Efrain~J. Ferrer and Vivian de~la Incera.
\newblock {"Magnetism in Dense Quark Matter"}.
\newblock {\em Lect.~Notes Phys.}, 871:397, 2013.
\newblock arXiv:1208.5179[nucl-th].

\bibitem{Landsteiner:2012kd}
Karl Landsteiner, Eugenio Megias, and Francisco Pena-Benitez.
\newblock {"Anomalous Transport from Kubo Formulae"}.
\newblock {\em Lect.~Notes Phys.}, 871:431, 2013.
\newblock arXiv:1207.5808[hep-th].

\bibitem{D'Hoker:2012ih}
Eric D'Hoker and Per Kraus.
\newblock {"Quantum Criticality via Magnetic Branes"}.
\newblock {\em Lect.~Notes Phys.}, 871:467, 2013.
\newblock arXiv:1208.1925[hep-th].

\bibitem{Bzdak:2012ia}
Adam Bzdak, Volker Koch, and Jinfeng Liao.
\newblock {"Charge-Dependent Correlations in Relativistic Heavy Ion Collisions
  and the Chiral Magnetic Effect"}.
\newblock {\em Lect.~Notes Phys.}, 871:501, 2013.
\newblock arXiv:1207.7327[nucl-th].

\bibitem{Albash:2012ht}
Tameem Albash, Clifford~V. Johnson, and Scott MacDonald.
\newblock {"Holography, Fractionalization and Magnetic Fields"}.
\newblock {\em Lect.~Notes Phys.}, 871:535, 2013.
\newblock arXiv:1207.1677[hep-th].

\bibitem{Gubankova:LNP}
E.~Gubankova, J.~Brill, M.~\^Cubrovi\^c, K.~Schalm, P.~Schijven, and J.~Zaanen.
\newblock {"Holographic description of strongly correlated electrons in
  external magnetic fields"}.
\newblock {\em Lect.~Notes Phys.}, 871:553, 2013.
\newblock arXiv:1304.3835[hep-th].

\bibitem{Bergman:2012na}
Oren Bergman, Johanna Erdmenger, and Gilad Lifschytz.
\newblock {"A Review of Magnetic Phenomena in Probe-Brane Holographic Matter"}.
\newblock {\em Lect.~Notes Phys.}, 871:589, 2013.
\newblock arXiv:1207. 5953[hep-th].

\end{thebibliography}

\end{document}